\documentclass[aps,prb,twocolumn,groupedaddress,superscriptaddress,amsfonts,amssymb,amsmath,longbibliography]{revtex4-2}
\usepackage{siunitx}
\usepackage{graphicx,float,xcolor}

\begin{document}


\title{Spin-selective resonant tunneling induced by Rashba spin-orbit interaction\\ in semiconductor nanowire}



\author{J. Paw\l{}owski}
\email[]{jaroslaw.pawlowski@pwr.edu.pl}
\affiliation{
	Department of Theoretical  Physics,
	Wroc\l{}aw University of Science and Technology, Wybrze\.{z}e Wyspia\'{n}skiego 27, 50-370 Wroc\l{}aw, Poland}
\author{G. Skowron}
\author{P. Szumniak}
\author{S. Bednarek}
\affiliation{
	Faculty of Physics and Applied Computer Science,
	AGH University of Science and Technology, Krak\'{o}w, Poland}


\date{\today}

\begin{abstract}
We consider a single electron confined within a quantum wire in a system of two electrostatically-induced QDs defined by nearby gates. The time-varying electric field, of single GHz frequency, perpendicular to the quantum wire, is used to induce the Rashba coupling and enable spin-dependent resonant tunneling of the electron between two adjacent potential wells with fidelity over $\SI{99.5}{\percent}$. This effect can be used for the high fidelity all-electrical electron-spin initialization or readout in the spin-based quantum computer. In contrast to other spin initialization methods, our technique can be performed adiabatically without 
increase in the energy of the electron. 
Our simulations are supported by a realistic self-consistent time-dependent Poisson-Schroedinger calculations.
\end{abstract}


\maketitle

\section{Introduction}
Fast and precise initialization of two-level quantum systems (or qubits)~\cite{1,2} is one of the main requirements for physical implementation of a quantum computation. At the start of computations qubits need to be precisely initialized.
Secondly, the quantum error correction schemes require a source of of readily available and
quickly initialized qubits. Spins of electrons in solids are attractive candidates for qubits due to their relatively long coherence times and potential scalability~\cite{sq1,sq2,sq3}. There are number of spin-initialization methods that differ in initialization time, fidelity and complexity of implementation. Two common approaches exploiting:  spin relaxation process and spin-resolved exchange of an electron with an external reservoir are rather too slow for a practical use. Among other more robust schemes, optical initialization via excited trion states~\cite{3,4} provides high fidelity exceeding $\SI{99}{\percent}$~\cite{4,5}. Whereas in gated quantum dots (QDs) the most common approach is to employ the Pauli spin-blockade mechanism~\cite{8,9,10}, however it requires an additional previously initialized spin (e.g. through relaxation).

Another family of spin initialization and manipulation methods exploit the electric field induced Rashba spin-orbit interaction (RSOI), that couples the electron spin with the momentum~\cite{r1,r2}. Modulation of this coupling allows to control the operation of spintronic devices \cite{jap1,jap2,all2}. Its sudden changes can be used to set an electron in motion in the spin-dependent direction~\cite{myold,moskal}, but also to initialize its spin all-electrically~\cite{ultraszybka,my2,pawel}. Several recent studies suggest that spin-orbit effect can be exploited for realization of spin-polarization via resonant tunneling without the need of an external magnetic fields neither application of ferromagnetic materials. Such spin-dependent electron tunneling through a semiconductor barrier is caused by the Rashba~\cite{11,12,13,14,15,all2} and/or the Dresselhaus~\cite{16,17,18,19,20,24,26,21,22} spin-orbit effect present at the barrier region.
Its sensitivity to spin-orbit coupling also enables to determine parameters of the spin-orbit interaction or build precision electric field sensors \cite{sen1,sen2,sen3}.
Unfortunately these propositions usually do not provide polarization fidelity better than $\SI{90}{\percent}$.
The spin filtering efficiency can be improved if the Pauli spin-blockade is also employed~\cite{23}, however this further complicates the spin initialization process.

Here we propose QD spin initialization technique which utilize the spin-selective resonant tunneling in a different way. The tunnelling between two adjacent QDs is induced by a time-varying RSOI while the confinement potential remains virtually unchanged. This allows to achieve high initialization fidelity.
%
Variable RSOI was also the basis for an efficient spin-initialization scheme presented in~\cite{ultraszybka,my2}. However, in that case the electron gained a significant amount of energy during the spin  initialization process which could not be easily relaxed, limiting the overall fidelity of the process. This work, however, shows that the spin-selective tunneling can be achieved through slow (adiabatic) changes of the RSOI combined with resonant tunneling through a barrier without unwanted energy transfer, resulting in high initialization fidelity.
One should note that  similarly high initialization fidelities  are also achievable much  faster, in non-adiabatic regime \cite{ ewgienij2,ewgienij1}, however, it requires much larger electric fields that drive the spin-orbit coupling.

\section{Theory}
\begin{figure}[t]
\centering
\includegraphics[width=0.45\textwidth]{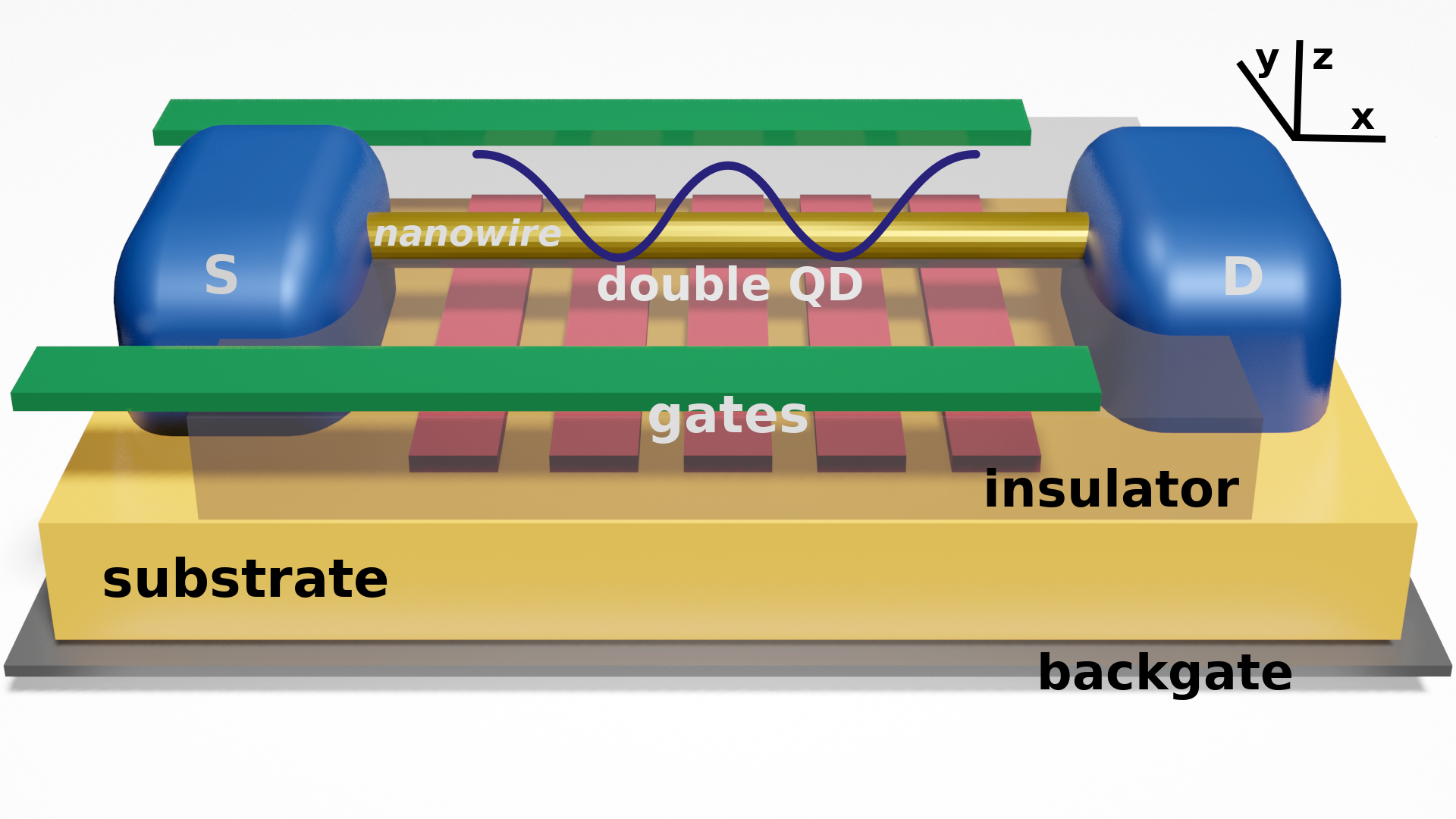}
\caption{\label{fig1}Schematic view of the proposed nanodevice. An array of bottom gates (red) is used to create the double QD confining potential within the $\text{InSb}$ nanowire connected at both ends to source (S) and drain (D) electrodes. Two lateral gates (green) are used to create an electric field used to induce the Rashba spin-orbit coupling. 
}
\end{figure}
Let us consider a single electron trapped in a semiconductor nanowire parallel to the $x$-axis as presented in Fig.~\ref{fig1}. The Hamiltonian of the spinfull system takes the form:
\begin{equation}\label{eq:hamiltonian_symbolic}
\mathbf{\hat{H}}=\hat{H}^0(x)\mathbf{1}_2+\mathbf{\hat{H}}_\text{so},
\end{equation}
where $\hat{H}^0(x)$ includes the kinetic and potential energy operators:
$\hat{H}^0(x)=-\frac{\hbar^2}{2m^\ast}\partial^2_x+V(x)$,
and $\mathbf{\hat{H}}_\text{so}$ describes the spin-orbit part. By $m^\ast$ we denote the electron effective mass. The wave function of the electron takes a 2-row spinor form
$\boldsymbol\Psi(x,t)=\left(\psi_\uparrow(x,t),\psi_\downarrow(x,t)\right)^T$.
The RSOI is generated by an external electric field, perpendicular to the quantum wire~\cite{dre}:
\begin{equation}\label{eq:hamiltonian_so}
\mathbf{\hat{H}}_\text{so}=(\beta_z\hat{\sigma}_y-\beta_y\hat{\sigma}_z)\hat{p}_x,
\end{equation}
where $\hat{\sigma}_y$, $\hat{\sigma}_z$ are the Pauli matrices, while coefficients $\beta_z$ and $\beta_y$ depend on the electric field components: $\beta_{z,y}=\alpha_\text{so}|e|E_{z,y}/\hbar$, with $\alpha_\text{so}$ being the RSOI coupling constant.
Let us now assume, that $E_y$ is the only non-zero component of the electric field. This effectively eliminates the first term in Eq.~\eqref{eq:hamiltonian_so}. For the constant field, the eigenproblem can be solved analytically. The corresponding ground state is doubly degenerate with respect to the spin, with the following wavefunctions:
\begin{equation}\label{eq:wavefunctions}
\mathbf{\Psi}_\uparrow(x)=\begin{pmatrix}1\\0\end{pmatrix}\varphi(x)e^{iqx},\quad\mathbf{\Psi}_\downarrow(x)=\begin{pmatrix}0\\1\end{pmatrix}\varphi(x)e^{-iqx},
\end{equation}
where $\varphi(x)$ is an eigenstate of the operator $\hat{H}^0(x)$: $\hat{H}^0\varphi(x)=\mathcal{E}\varphi(x)$, and
$q=m^\ast\beta_y/\hbar$.
Notice, that the wavevectors in both spin states have opposed sign. Applying the Hamiltonian \eqref{eq:hamiltonian_symbolic} to $\mathbf{\Psi}_{\uparrow/\downarrow}$ yields
\begin{equation}\label{eq:eigeneqsimple}
\mathbf{\hat{H}}\mathbf{\Psi}_{\uparrow/\downarrow}(x)=\left(\mathcal{E}-\mathcal{E}_q\right)\mathbf{\Psi}_{\uparrow/\downarrow}(x),
\end{equation}
with the spin-orbit energy $\mathcal{E}_q=\frac{\hbar^2q^2}{2m^\ast}$.

Let us now consider nontrivial dynamics of the system induced by a time varying electric field $E_y(t)$.
%
We assume that the shape of potential $V(x)$ does not change over time, but the spin-orbit coupling changes adiabatically (slowly enough). In such conditions, 
electron follows the ground-state during time evolution,
i.e. the wavefunctions are still of the form given by \eqref{eq:wavefunctions} but with the wave vector changing over time as:
\begin{equation}
q(t)=m^\ast\alpha_\text{so}|e|E_y(t)/\hbar^2.
\end{equation}
To simplify calculations we assume the following form of the wavefunction $\boldsymbol\Psi(x,t)$ components:
\begin{equation}\label{eq:wavefunctionassumed}
\psi_\sigma(x,t)=\Phi(x,t)e^{i\sigma{}q(t)x},
\end{equation}
with $\sigma=+1$ ($-1$) for spin-up (spin-down). Notice that $|\psi_\sigma(x,t)|^2=|\Phi(x,t)|^2$, therefore $\Phi(x,t)$ has the full information about the electron position. 
Let us check how it changes over time by inserting into
time-dependent Schr\"odinger equation with the Hamiltonian \eqref{eq:hamiltonian_symbolic}:
\begin{equation}
i\hbar\partial_t\psi_\sigma(x,t)=\left(H^0(x)-\mathcal{E}_q(t)\right)\Phi(x,t)e^{i\sigma{}q(t)x}.
\end{equation}
By calculating the time-derivative
\begin{equation}
i\hbar\partial_t\psi_\sigma(x,t)=\left(i\hbar\partial_t\Phi(x,t)-\sigma\hbar{}\dot{q}(t)x\Phi(x,t)\right)e^{i\sigma{}q(t)x},
\end{equation}
one can obtain a time evolution of $\Phi(x,t)$ in the form of the time-dependent Schr\"odinger equation
$i\hbar\dot{\Phi}(x,t)={\tilde{H}(x,t)}\Phi(x,t)$, albeit with the modified Hamiltonian $\tilde{H}$:
\begin{equation}\label{eq:ham_tilde}
\tilde{H}(x,t)=\hat{H}^0(x)-\mathcal{E}_q(t)+\sigma\hbar\dot{q}(t)x.
\end{equation}
It differs from $\hat{H}^0$, primarily by an additional term proportional to the speed of change of the spin-orbit coupling. The sign of this term depends on the electron spin-state (up or down). Now, the two groundstates of $\tilde{H}$ with opposite spins are no longer degenerate and their corresponding eigenvalues are shifted with respect to each other \cite{kramers}. 
%
We can further exploit this behavior in resonant tunneling through a potential barrier. 

\section{Resonant transmission}
\begin{figure}[b]
\centering
\includegraphics[width=0.48\textwidth]{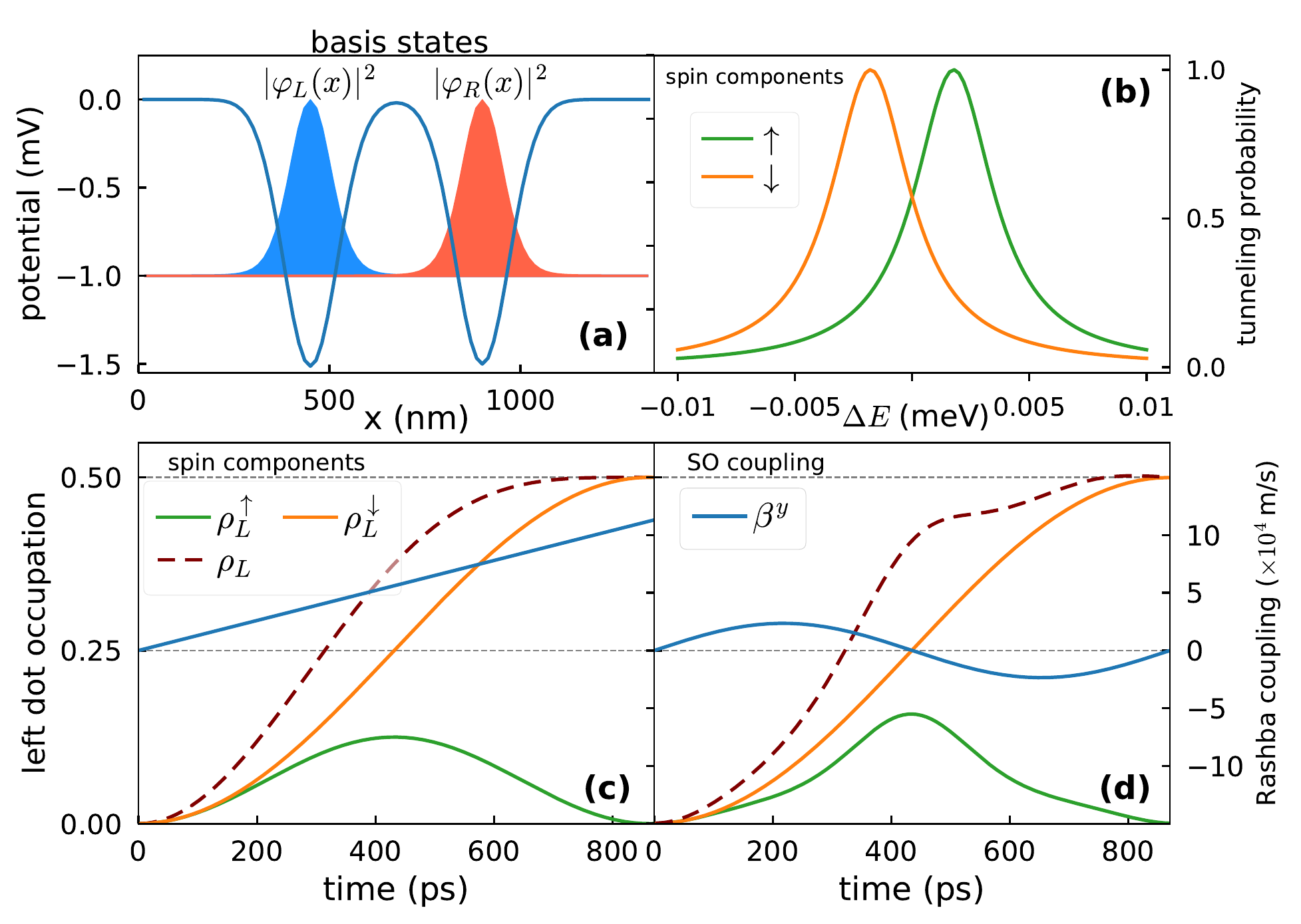}
\caption{\label{fig2}(a) Confinement potential (blue) of the double QD structure, and the two assumed basis state localized in the left and right dot each: ground-state-wavefunction in the left $\varphi_\text{L}(x)$ (blue), and in the right dot $\varphi_\text{R}(x)$ (red). (b) Maximum probability of the electron transition between the dots for spin-up (green) or spin-down (orange) orientation. Spin-selective resonant tunneling for the linear (c) and sinusoidal (d) time-dependence profile of the Rashba coupling: shown by the left-dot-occupancy for spin-up (green) and spin-down (orange) wavefunction part, and their sum $\rho_L$ (electron is initially in the right dot with the spin oriented along $x$-axis).}
\end{figure}
Let us assume that confinement potential has the form of a double QD as presented in Fig.~\ref{fig2}(a). If the bottom of the right dot is lower than the left and barrier sufficiently high, an electron in the ground-state is localized exclusively in the right dot. If we now align bottoms of both dot potentials (the ground-state energies for both dots equalize) the electron starts tunneling through the barrier to the left dot. After the transition completes, it returns to the right dot. Then the whole process repeats and  continues cyclically resulting in oscillations of occupation. This process can be described using a two state basis~\cite{feynman}, where the wavefunction is expressed as a linear combination of two basis functions $\varphi_\text{L,R}(x)$ located in the left and right dot---see Fig.~\ref{fig2}(a). The basis functions are ground-states of the system with slightly deepened left or right dot respectively. We also adjust the width and height of the barrier, so that the basis states slightly overlap, and they are not perfectly orthogonal. This directly determines the tunneling speed.

To examine conditions that enable tunneling between the dots we calculate the  matrix elements of the Hamiltonian (\ref{eq:ham_tilde}) in both basis states and require them to be equal:
\begin{equation}\label{eq:expval_equal}
\tilde{H}_\text{RR}-\tilde{H}_\text{LL}=H^0_\text{RR}-H^0_\text{LL}+\sigma\hbar\dot{q}(t)(x_\text{RR}-x_\text{LL})=0,
\end{equation}
where $x_\text{LL}$=$\langle \varphi_\text{L}|x| \varphi_\text{L}\rangle$ ($x_\text{RR}=\langle \varphi_\text{R}|x| \varphi_\text{R}\rangle$) are the expectation values of the electron position in the left (right)-dot-ground-state. Their difference is the distance between minima of the QDs. If the spin-orbit coupling changes over time, the last term in (Eq. \ref{eq:expval_equal}) is nonzero: $\dot{q}(t)\neq{}0$. 
In order to ensure the resonant tunneling process, one have to shift the bottoms of the QDs by a spin dependent energy $\Delta{}E$, which we describe below.
Since $\sigma$ is $1$ for spin-up and $-1$ for spin-down, if we are to allow for resonant tunneling, we have to shift the bottoms of dots by a different energy (sign) depending on spin orientation. This way we select for which spin orientation the resonance occurs. 

Let us now write the electron wave function as a linear combination of the basis states \cite{feynman}:
$\Phi(x,t)=C_\text{L}(t)\varphi_\text{L}(x)+C_\text{R}(t)\varphi_\text{R}(x)$.
We put this linear combination into the Schr\"odinger equation for $\Phi(x,t)$ (see Eq. \ref{eq:ham_tilde}) and obtain a set of equations for time-evolution of both expansion coefficients. Its analytic solutions can be easily obtained for a linear $q(t)$. In such a case, the time-derivative $\dot{q}(t)$ is constant (the last term in \eqref{eq:expval_equal}), being a SOI-induced energy difference between the right and left dot energy levels. Let us denote this difference as $\hbar\gamma$:
\begin{equation}\label{eq:energy_diff}
\Delta{}E=\sigma\hbar\dot{q}(t)(x_\text{RR}-x_\text{LL})=\sigma\hbar\gamma.
\end{equation}

In order to tunnel resonantly the spin-down electron (initially located in the right dot) to the left dot, we have to shift the bottom of the left dot by $-\hbar\gamma$ at the beginning of the process. After solving the equations for $C_\text{L}(t)$ and $C_\text{R}(t)$ we calculate the squared modulus $|C_\text{L}(t)|^2$ and $|C_\text{R}(t)|^2$ to obtain probabilities that the spin-down-electron occupies the left and the right QDs, respectively. These probabilities exhibit occupation oscillations:
\begin{equation}
P^\downarrow_\text{L}(t)=\sin^2(\omega_0{}t),\quad
P^\downarrow_\text{R}(t)=\cos^2(\omega_0{}t).
\end{equation}
The frequency depends on the off-diagonal matrix elements of the Hamiltonian $H_\text{LR}~$\cite{tilde}: $\hbar\omega_0=|H_\text{LR}|$. At the beginning we start with $P^\downarrow_\text{L}(t=0)=0$ and after half of the occupation oscillation period (which equals $T=\pi/\omega_0$), at $t_1=T/2=\pi/(2\omega_0)$ we get $P^\downarrow_\text{L}(t_1)=1$.
The $\Delta{}E$ differs in sign for spin-up and spin-down state. 
This implies that in the resonance conditions for spin-down (spin-up is out of resonance) one gets 
$\tilde{H}_\text{RR}-\tilde{H}_\text{LL}=2\hbar\gamma$. It is worth to note that, the spin-up-electron tunneling also occurs, however in the off-resonant fashion with a lower amplitude and  higher frequency $\omega_1=\sqrt{\omega_0^2+\gamma^2}$. The resonance conditions of the electron tunneling for both spin orientations are presented in Fig.~\ref{fig2}(b). The probabilities for the spin-up electron occupying the left and the right dot are given by
\begin{gather}
P^\uparrow_\text{L}(t)=\frac{\omega_0^2}{\omega_1^2}\sin^2(\omega_1t),\\
P^\uparrow_\text{R}(t)=\cos^2(\omega_1t)+\frac{\gamma^2}{\omega_1^2}\sin^2(\omega_1t).
\end{gather}
The angular frequency $\omega_1$ depends on $\gamma$, which in turn depends on the speed of change of the SOI. We can adjust it so that $\omega_1=2\omega_0$, which is achieved for $\gamma=\sqrt{3}\omega_0$. In such a case, while the spin-down electron passes completely from the right to the left dot -- what happens at time $t_1$ -- the spin-up electron passes to the left dot and returns back. 

This effect is used to separate the electron-spin. 
Let us assume that initially the electron has spin oriented along the $x$ axis, and is located in the right dot. We set bottoms of both dot-potentials, so that the resonance occurs for spin-down. We demand $\gamma=\sqrt{3}\omega_0$, thus after time $t_1$ the lower spinor component passes to the left dot, while the upper part returns to the right dot: $\boldsymbol\Psi(x,t_1)=\left(\psi_\text{R}(x,t_1), \psi_\text{L}(x,t_1)\right)^T$. The spin separation process is presented in Fig.~\ref{fig2}(c) with green (orange) curve showing electron density for spin-up $\rho_L^\uparrow$ (down $\rho_L^\downarrow$) component calculated over the left dot.
 
\section{Sinusoidal-variable spin-orbit coupling}
It turns out that we can achieve a similar separation of spin components by using a sinusoidal-variable spin-orbit coupling, for which $q(t)=q_0\sin(\omega{}t)$. In this case the Rashba-induced energy offset is $\Delta{}E=\sigma\hbar{}q_0\omega\cos(\omega{}t)(x_\text{RR}-x_\text{LL})$, and we have to shift the bottom of the left dot by such an energy in order to achieve the resonance for spin-down. The spin-down wave function component behaves as in the case of linear in time increase of RSOI, performing characteristic Rabi oscillations with frequency $\Omega=2\omega_0$. 

At the same time, the spin-up wave function component tunnels through the barrier between dots in the off-resonant fashion. To obtain off-resonant tunneling, which is twice as fast as the on-resonant one, we have to assume the driving frequency to be $\omega=4\omega_0$, and amplitude $q_0$ chosen according to the relation $q_0=j_1/(2(x_\text{RR}-x_\text{LL}))$, with $j_1$ being the 1\textsuperscript{st} zero of the $J_0(x)$ Bessel function. Then the spin-up electron behaves in a way similar to the case with the linear $q(t)$ dependence. It passes partially to the left dot and then returns to the right one at time $t_1$. Since it is not possible to get exact analytic solutions for the expansions coefficients $C_{L,R}$ (albeit we have found some asymptotic solutions), we find them numerically.
The results are presented in Fig.\ref{fig2}(d).
The probability of finding the spin-down electron in the left dot (orange curve) grows from $0$ to $0.5$ after $t_1=T/2=\pi/(2\omega_0)$, which means that the entire spin-down electron density is now located in the left dot
(initially the electron was set in equal superposition of up and down spin states). On the other hand, the probability of finding the spin-up electron in the left dot (green) grows from $0$ to about $0.15$ and then falls back to zero. Their sum (dashed brown) constitutes the probability of finding the electron in the left dot regardless of its spin. At time $t_1$ half of the wavepacket (spin-down) tunnels entirely to the left dot and the other half (spin-up) stays in the right dot.

\section{Spin initialization}
\begin{figure}[tb]
\centering
\includegraphics[width=0.48\textwidth]{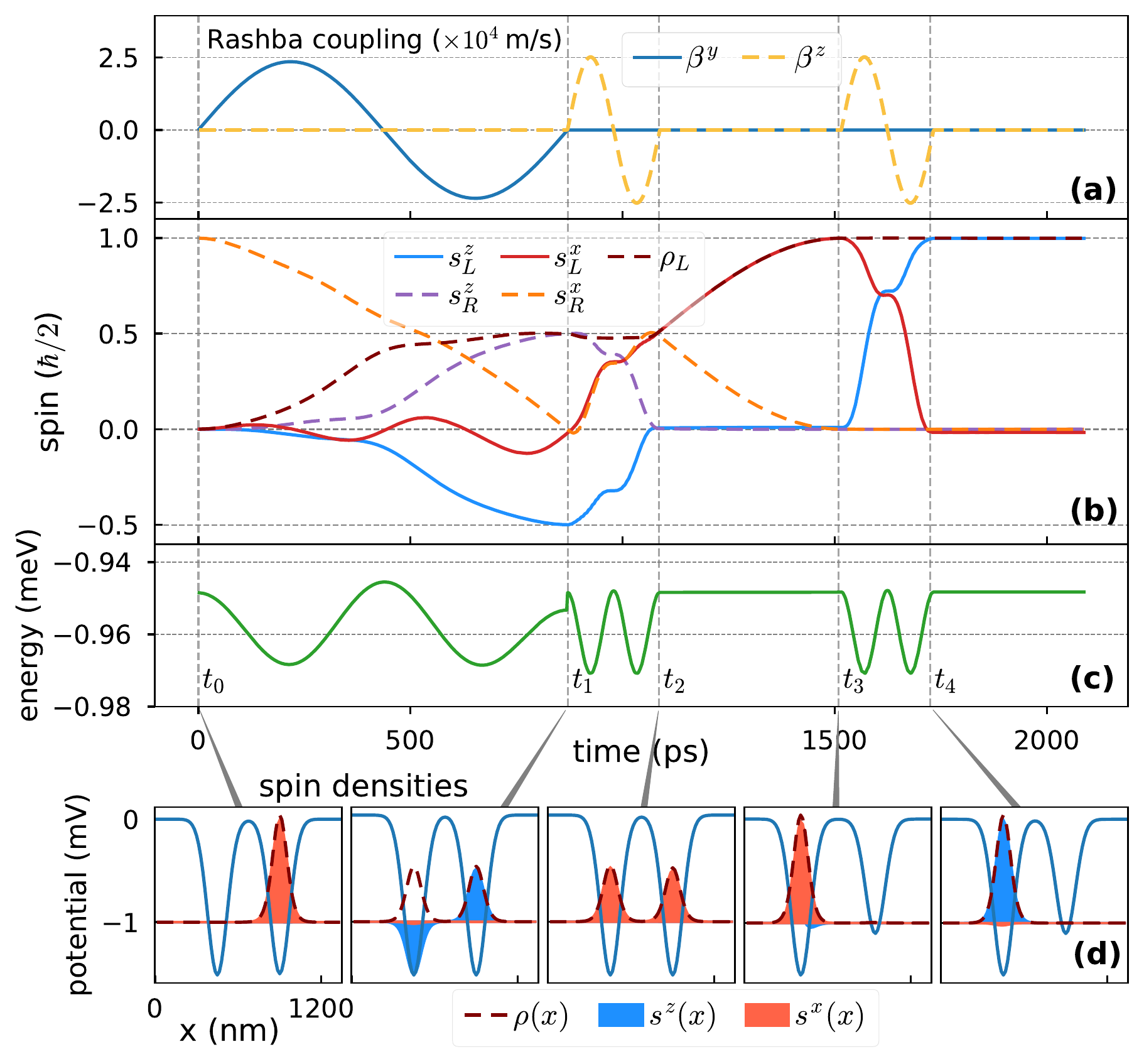}
\caption{\label{fig3} Electron spin initialization scheme including the spin-selective resonant tunneling followed by the spin rotation and final merging into a single initialized spin. (a) Time-dependence of the Rashba coupling: $\beta_y$ inducing the resonant tunneling (blue), and $\beta_z$ responsible for the spin rotations (yellow dashed). (b) Electron spin $x$ and $z$-component calculated separately: over the left $s^{x,z}_L$, and right dot $s^{x,z}_R$. (c) The total energy during the entire process. (d) The spatial spin density (red for the $x$ and blue for the $z$-component) and the total electron density (dashed brown) at the subsequent stages $t_{0..4}$ of the initialization.
}
\end{figure}

Here we show how spin-selective resonant tunneling phenomena can be used to initialize electron spin qubits hosted in gated-nanowire QDs (see Fig.~\ref{fig3}). This is of particular interest for spin based quantum computing implementations.
The spin initialization process is illustrated in Fig.~\ref{fig3}.
At the beginning, the electron is confined in the right dot with spin being parallel to the $x$-axis ($S_R^x=1$).
In the final state at $t_4$ time the electron occupies the left dot with spin-up (last frame in Fig.~\ref{fig3}(d)). In Fig.~\ref{fig3}(d) the dashed brown curve depicts the electron density expressed as $\rho(x,t)=\Phi^\dagger(x,t)\Phi(x,t)$
while the blue shaded area depicts the spin $z$-projection density calculated according to
\begin{equation}
\begin{split}
s^z(x,t) &= \Phi^\dagger(x,t)\sigma_z\Phi(x,t)\\&=\psi_\uparrow^\ast(x,t)\psi_\uparrow(x,t)-\psi_\downarrow^\ast(x,t)\psi_\downarrow(x,t).
\end{split}
\end{equation}
Spin density $s^x(x,t)$ is defined analogously. Spin density $s^z$ is positive (spin-up) or negative (spin-down) depending on which spin orientation dominates in a particular area. In the first frame of Fig.~\ref{fig3}(d) spin density $s^z(x)$ is zero everywhere (the blue area is absent), which indicates that the spin state is an equally-weighted linear combination of spin-up and spin-down. In the last frame spin density $s^z(x)$ and electron density $\rho(x)$ overlap which means the electron is in the spin-up state with 100\% probability.

The spin initialization requires a sequence of four steps. The first one involves the already discussed spin-selective resonant tunneling in which the spin-down component ends up in the left dot. This is achieved by using a sine-shaped pulse of the Rashba coupling: $\beta_y^\text{so}=\beta_1\sin(4\omega_0{}t)$ with an appropriate amplitude.
The simulations were performed for a nanowire made of InSb, a material with strong RSOI coupling $\alpha_\text{so}=\SI{523}{\angstrom^2}$\cite{winkler}.  
The $\beta_1$ value has to be tuned to meet the resonance conditions, as presented in Fig.~\ref{fig2}(b).
The required value of $\beta_1$ is achieved for an electric field of $\SI{2.7e6}{\volt\per\metre}$, which is easily achievable in semiconductor nanostructures. If we assume material constants for InAs ($\alpha_\text{so}=\SI{117.1}{\angstrom^2}$) the required electric field amplitude is higher and equals about $\SI{8.65e6}{\volt\per\metre}$. In GaAs the Rashba coupling is about 100 times weaker ($\alpha_\text{so}=\SI{5.2}{\angstrom^2}$) and the required electric field of $\SI{6.67e7}{\volt\per\metre}$ is still feasible, but can make more difficult to realize the device.
Note that all this mentioned electric field amplitudes assume certain values of the Rashba coupling present in the structure. However, realistic values of RSOI couplings are structure- and device-dependent, therefore required values of the electric field will have to be founded experimentally for a given nanodevice setup.
The first initialization step ends at $t_1 = T/2 = \pi/(2\omega_0)$. The electron remains in the state depicted in the 2\textsuperscript{nd} frame of Fig.~\ref{fig3}(d). The wave function is divided into two components. The first one, with spin-down, now occupies the left dot, while the second one, spin-up, remains in the right dot. 

The second step involves spin rotations about the $y$-axis. Spin components located in both QDs are rotated by an angle of $\pi/2$ but in opposite directions. This effectively aligns spin in both dots in parallel to the $x$-axis. The rotation is achieved by introducing a time-varying spin-orbit coupling according to the formula $\beta_z^\text{so}=\beta_2\sin(16\omega_0{}t)$ and by slightly moving the minima of both potential wells in opposite directions \cite{myold2}, by a distance $\Delta{}x_{1,2}=\pm{}A[1-\cos(16\omega_0{}t)]$, with assumed $A=\SI{95}{nm}$.
While each spinor components move apart in opposite directions and then get back, spin in the left and the right QD rotates by an angle $\pi/2$ and $-\pi/2$, respectively (the RSOI amplitude $\beta_2$ must be properly tuned). This step ends at $t_2=t_1+T/8=\pi/(8\omega_0)$, shown in the 3\textsuperscript{rd} frame of Fig.~\ref{fig3}(d). Spins in both QDs are now parallel to the $x$-axis. 

During the third step, the electron wave function parts from both QDs have to be be merged into a single one. 
This can be achieved by allowing for tunneling into the left QD, which is now faster and at $t_3=t_2+T/4$ the electron occupies the left dot with 100\% probability. At this moment we block tunneling in the reverse direction by increasing the bottom of the right dot potential (4\textsuperscript{th} frame of Fig. \ref{fig3}(d)). The last (fourth) step involves an additional spin rotation by an angle $\pi/2$ to align the spin with the $z$-axis. We proceed exactly as in the 2\textsuperscript{nd} step. The spin is now initialized (oriented along the $z$-axis) and the electron occupies fully the left QD. 

The entire procedure takes $t_\text{tot}=\pi/\omega_0$ time and depends on the tunneling speed which can be controlled by tuning the shape of the potential barrier between the QDs. The process is characterized by exceptionally high fidelity (probability of spin-up) of about 99.9\% for the $\hbar\omega_0=\SI{0.0012}{meV}$, which gives the total initialization time $t_\text{tot}=\SI{1726}{ps}$. 
One of the main advantages of the proposed spin initialization technique is shown in Fig. \ref{fig3}(c). The total energy, initially equal $E_\text{start}=\SI{-0.94862}{meV}$, changes during operation of the nanodevice but at the end reaches the value $E_\text{end}=\SI{-0.94843}{meV}$, which is nearly identical with the initial one. The time of initialization (already very fast) can be further shortened but for the price of fidelity. If we quadruple the tunneling speed ($\hbar\omega_0=\SI{0.0048}{meV}$) the initialization takes only $\SI{431}{ps}$ but the fidelity drops to $99.5\%$, with a higher change in the total energy equal $\SI{0.0012}{meV}$, which is still very good result.
It is worth to underline that the reported nonideal fidelities are due to the limitations of the method itself, not the loss of  the qubit’s coherence due to decoherence.

Fidelities above $99.5\%$  are also achievable much faster, in non-adiabatic regime \cite{ewgienij2, muga, ewgienij1}, leading to initialization times of 1.5 ns reported for GaAs \cite{ewgienij2}, and correspondingly shorter for materials with stronger  spin-orbit coupling. However, such non-adiabatic schemes require 1-2 orders of magnitude larger electric fields that drive the Rashba coupling, which can make more difficult to realize the device, because we have to make sure that the carrier does not tunnel off the QD under such strong electric field.

\section{Summary}
We showed that slow (adiabatic) changes of the Rashba spin-orbit coupling induce a spin-dependent energy shift, for which the resonant tunneling between two spatially separated (yet adjacent) QDs occurs. If we tune the rate of changes of the spin-orbit coupling appropriately, it is possible to ensure that the spin-down electron tunnels in the resonant fashion, while the spin-up electron in off-resonant one but twice as fast. 
We showed that this effect can be used to initialize or read out electron spin, fulfilling crucial requirements for realizing spin-based quantum computer. We have performed calculation for realistic system---InSb gated nanowire and showed that spin initialization fidelity, due to adiabatic nature of the proposed method, can reach as high as $99.5\%$ for an initialization time of $\SI{400}{ps}$. 
This is two orders of magnitude faster than the spin coherence time in InSb material of  $\SI{34}{ns}$ \cite{nadj}, reported for  nanodevice of structure similar to ours.
The entire procedure is all-electrical and all-semiconductor \cite{all,all2} with voltages oscillating with single GHz frequencies and amplitudes below $\SI{1}{V}$ used to induce the Rashba coupling. This all  makes our proposal particularly suitable for scalability purposes.

\section{Acknowledgements}
PS acknowledges support within POIR.04.04.00-00-5CE6/18 project carried out within the HOMING programme of the Foundation for Polish Science co-financed by the European Union under the European Regional Development Fund. Part of calculations have been carried out using resources provided by Wroclaw Centre for Networking and Supercomputing (wcss.pl), grant No.~520.

\bibliography{thebibliography}

\begin{thebibliography}{50}
\bibitem [{dre()}]{dre}%
  \BibitemOpen
  \href@noop {} {}\bibinfo {note} {We assume that the quantum wire is grown
  along the $[1,1,1]$ crystallographic direction, thus the Dresselhaus
  interaction originating from the (zinc-blende) structure asymmetry is
  negligible \cite{winkler}.}\BibitemShut {Stop}%
\bibitem [{kra()}]{kramers}%
  \BibitemOpen
  \href@noop {} {}\bibinfo {note} {This could be expected since time-varying
  SOI break the time-reversal symmetry which lifts the Kramers
  spin-degeneracy.}\BibitemShut {Stop}%
\bibitem [{til()}]{tilde}%
  \BibitemOpen
  \href@noop {} {}\bibinfo {note} {We introduce
  $H_\text{LR}=(\tilde{H}_\text{LR}-P_\text{LR}\tilde{H}_\text{RR})/(1-|P_\text{LR}|^2)$
  due to nonorthogonality (overlapping) of the basis states, leading to nonzero
  $P_\text{LR}=\langle\varphi_L|\varphi_R\rangle$ element. Nonzero overlapping
  enables tunneling process itself.}\BibitemShut {Stop}%
\end{thebibliography}

\end{document}